# Application of Artificial Neural Network in the Control and Optimization of Distillation Tower

Chunli Li[1], Chunyu Wang [2],*

[1]School of Chemical Engineering and Technology, Hebei University of Technology, Tianjin 300130, China;

[2]School of Chemical Engineering, Hebei University of Technology, Shanghai 200237, China;

* Corresponding author:

**Abstract**

Distillation is a unit operation with multiple input parameters and multiple output parameters. It is characterized by multiple variables, coupling between input parameters, and non-linear relationship with output parameters. Therefore, it is very difficult to use traditional methods to control and optimize the distillation column. Artificial Neural Network (ANN) uses the interconnection between a large number of neurons to establish the functional relationship between input and output, thereby achieving the approximation of any non-linear mapping. ANN is used for the control and optimization of distillation tower, with short response time, good dynamic performance, strong robustness, and strong ability to adapt to changes in the control environment. This article will mainly introduce the research progress of ANN and its application in the modeling, control and optimization of distillation towers.

## 1 Introduction

Distillation is a mass transfer separation process widely used in petrochemical, organic chemical and pharmaceutical production. It is characterized by large energy consumption, many variables, and complex constraints and difficult to control [1, 2]. If the distillation tower is well controlled, not only can the product quality and recovery rate be improved, but it is also beneficial to environmental protection and energy conservation. The rectification column is a complex nonlinear system, usually with dozens or even hundreds of trays, so the model order is relatively high, which is not convenient for theoretical analysis and real-time calculation. Therefore, it is required to establish a simplified model that can reflect the static and dynamic characteristics of the distillation column [3]. On the other hand, the conventional control method is difficult to achieve fast tracking and optimal control of the rectifying tower [4]. The development of advanced modeling and control methods has become a research hotspot.

Artificial Neural Network (ANN) is an information processing system based on imitating the structure and function of the brain's neural network. Using the interconnection between a large number of neurons to form a network system capable of complex calculations, it actually describes a functional relationship between network input and output [5]. The artificial neural network can achieve the approximation of any nonlinear mapping through learning, and it can be applied to the identification and modeling of nonlinear systems without being restricted by nonlinear models. It also has the ability to adapt, through learning and training to find out the internal connection between the provided data and output to make the problem answered, instead of relying on prior knowledge and rules to solve the problem [6]. In addition, ANN is highly fault-tolerant. Even if part of the system suffers losses, it can restore the original information without affecting the overall activities. For example, Shi et al. [7] proposed an affine invariance method to generate dense correspondence between uncalibrated wide baseline images. To this end, a reliable sparse matching strategy is proposed, which starts with affine invariant feature extraction and matching, and then uses these initial matches as spatial priors to generate more sparse matches. The matching propagation from the sparse feature points to its neighboring pixels is carried out in the way of region growth in the affine invariant framework. The unmatched points are processed by the low-rank matrix recovery technique. The experimental results showed that in the presence of large affine deformation, the proposed method has a significant improvement over the existing ones.

The unique structure and information processing method of ANN makes it have obvious advantages in many aspects and has a wide range of applications. The main application areas are signal processing [8,9], image processing [10-12], intelligent driving [13-15], robot control, automatic control of power systems, health care, medical treatment, chemical process control and optimization, etc. [16-18].

This article will mainly introduce the research progress of ANN and its application in the modeling and optimization of distillation column.

**2. Research progress of artificial neural networks**

It has been more than 70 years since McCulloch and Pitts [19] proposed the neuron biology model (M-P model) in 1943. The first 20 years are the initial stage, focusing on the proposal of the network model and the determination of the learning algorithm. Hebb learning rules [20] is still a basic principle of neural network learning algorithms; Rosenblatt et al. [21] and Widrow et al. [22] respectively proposed the perceptron model and the adaptive (Adaline) linear component model. Among them, the perceptron is the first ANN that is physically constructed and has the ability to learn. However, Minsky et al. [23] pointed out that Rosenblatt's single-layer perceptron can only learn linearly separable patterns, and cannot handle linearly inseparable problems such as xor. In the next 20 years, due to the limitation of the speed of computers, the research of neural networks entered a period of stagnation, but some important results have also been achieved. For example, a unified theory was proposed by closely combining mind and brain [24]. Subsequently, the development of neural network system theory entered a golden age. Hopfield [25] proposed a Hopfield network model that mimics the human brain, which is dynamic and may be used to solve complex problems. Rumelhart et al. [26] proposed back-propagation neural network (BPNN) to compensate for the shortcomings of multi-layer neural networks, but BPNN also has some problems, such as slow convergence and difficulty in converging for large sample data, prone to local minimums. Moody and Darken [27] proposed the Radial Basis Neural Network (RBFNN), which is an abstraction and simplification of the human brain neural network system The RBF neural network is a three-layer feedforward neural network, which contains an input layer, a hidden layer, and an output layer. The transformation from input nodes to hidden layer nodes is nonlinear, and the transformation from hidden layer nodes to output nodes is It is linear, so it can approximate any continuous function with arbitrary precision, which is very suitable for nonlinear dynamic modeling [28]. Compared with BPNN, in addition to superior clustering and classification capabilities,

RBFNN also has better generalization capabilities and higher approximation accuracy [29]. Due to its simple learning algorithm and network structure, RBFNN has fast convergence and can uniformly approximate any continuous function to achieve the expected accuracy. Therefore, RBFNN is particularly suitable for the control of nonlinear and time-varying dynamic systems, but the uncertainty and parameter changes require additional attention [30]. In 1998, based on the convolution and pooling network structure, Lecun [31] proposed the original model of Convolutional Neural Network (CNN), LeNet-5. Chen et al. [32] proposed a dynamic optimization technology combined with a general dynamic matrix control algorithm neural network model, which optimizes the calculation performance and greatly reduces the calculation time. In 2006, Hinton et al. [33] proposed Deep Belief Network (DBN) and described a method by which high-dimensional data can be converted into low-dimensional code with a small central layer by training a multilayer neural network. The network is used to reconstruct high-dimensional input vectors [34], which has triggered a boom in the application of deep learning, especially in the fields of image, video and signal processing and prediction. Gao et al. [35] proposed a distributed mean-field-type filtering(DMF) framework to handle those noisy, partial-observed, and high-dimensional. The approach iterates through four operations: sampling, prediction, decomposition, and correction. DMF was implemented in aircraft and vehicle tracking scenarios, and the results showed that it is superior to the traditional non-average field filter. Wang et al. [36] established a short-term sales forecast model for e-commerce accounting systems based on historical e-commerce sales data and portal product link clicks. It uses the AdaBoost idea to aggregate the prediction results of multiple traditional BP neural networks, which has a significant improvement in accuracy compared to traditional BP networks. Martinez-Morales et al. [37] proposed an MLP-MOACO model by optimizing MLP parameters through a multi-objective ant colony optimization algorithm, and it was applied to calculate the correlation coefficient of engine pollutants and estimate engine exhaust emissions. Gao et al. [38] proposed a game theory framework to train generative models, in which the unknown data distribution is learned by dynamically optimizing the worst-case loss measured using the Wasserstein metric. NG et al. [39] proposed an improved back-propagation algorithm GBP, which modifies the partial derivative of the activation function to increase the error signal of the back-propagation and normalizes the learning rate of the algorithm to improve and speed up the convergence speed.

**2 Application research progress of artificial neural network in distillation process**

Since it can approach any function theoretically, and has a strong nonlinear mapping ability, the artificial neural network can be used to simulate and predict the complex reactive distillation process well. Artificial neural networks commonly used in distillation systems mainly include Hopfield network model, BPNN and RBFNN.

Ramchandran et al. [40] applied neural networks to model the steady-state inverse of a process which is then combined with a simple reference system synthesis to create a multivariable controller. The control strategy is used to dynamic simulations of two methanol-water distillation columns. A steady-state process simulation package was used to generate all the neural network training data. An efficient training algorithm was used to train the networks. The controllers show robust performance, and performed better than proportional-integral (PI) controllers with feedforward features.

Zhao et al. [41] used batch-by-batch processing and non-monotonic linear search methods to improve the standard BP algorithm and applied it to the simulation of the operation process of the methyl acetate catalytic distillation hydrolysis tower. A 3×6×1 network is used to learn 130 sets of sample data in this process, and 26 sets of samples are used for detection, which effectively avoids local minimums and converges faster. Wang et al. [42] proposed an improved BP algorithm by combining genetic algorithm and artificial neural network. A model was constructed based on the simulation results with ASPEN PLUS. The improved BP algorithm is applied to the simulation and optimization of the separation process of C5 component. The results showed that the relative error of the reboiler load is 3.70%, and the relative errors of the two sideline liquid products of the main tower (SB and SC) are 1.04% and 1.60%, respectively, which proved that the ANN algorithm has high calculation accuracy and high solving efficiency, and can effectively find the optimal operating conditions.

Shi et al. [43] used the RBF neural network in a reactive distillation tower for preparing dimethyl carbonate (DMC) by transesterification of propylene carbonate (PC) and methanol, and established the relationship between the conversion rate and the operating conditions such as pressure, reflux ratio, feed mole. After optimization, the suitable process conditions were obtained, and the conversion rate of propylene carbonate can reach 45.79%. Liu et al. [44] proposed a method for control of distillation prosses based on BP neural network, and applied it to the temperature control process of the isopropanol refining tower. The result showed that this method has the characteristics of short response time and high control accuracy.

Ma [45] used an artificial neural network (BPNN) model to simulate the distillation process of anisole. Firstly, the ASPEN software was used to perform traditional distillation simulation, and a training sample set was obtained from the simulation results and experimental data; then an artificial neural network is established by setting the input variables ( feed composition, feed flow rate, reflux ratio, number of feed plates, tower top pressure, tower top temperature) and the objective functions (the bottom temperature of the tower and the composition of top product). The optimal structure of the BPNN is selected as a 6×11×2 structure with a hidden layer transfer function of logsig. Comparing the simulation results with the experimental data, the average relative error of the top product composition is 0.3283%, which is significantly better than the ASPEN simulation result.

Konakom et al. [46] used neural network-based model predictive control (NNMPC) for conducting predefined optimal policy tracking determined by dynamic optimization strategy of a batch reactive distillation tower. Multi-layer feedforward neural network model and estimator are developed and used in the model predictive control algorithm. The results showed that the NNMPC provides satisfactory control performance than the P controller does in all cases.

Sharma [47] and others used three different control strategies: conventional PID control, model predictive control (MPC) and neural network predictive control (NNPC) to control the reactive distillation tower for the synthesis of tert-amyl methyl ether.

**Reference**

[1] Rév E, Emtir M, Szitkai Z, Mizsey P, Fonyó Z,. Energy savings of integrated and coupled distillation systems, Computers and Chemical Engineering, 25, 119-140, 2001.

[2] Jiang H , Xi Z , Rahman A A , Zhang X. Prediction of output power with artificial neural network using extended datasets for Stirling engines[J]. Applied Energy, 2020, 271:115123.

[3] Dong J, Qian J, Sun Y. Distillation tower control strategy and control structure, Refining and chemical industry automation, 1992, 5:22-25

[4] Wang H, Mo R. Review of neural network algorithm and its application in reactive distillation, Asian Journal of Chemical Sciences, 2021: 20-29.

[5] Zhao N, Lu J. Review of Neural Network Algorithm and its Application in Temperature Control of Distillation Tower, Journal of Engineering Research and Reports, 2021,

[6] Qi Y, Zheng Z.  Neural Network Algorithm and Its Application in Supercritical Extraction ProcessAsian, Journal of Chemical Sciences, 9(1): 19-28, 2021

[7] Shi F, Gao J, Huang X. An affine invariant approach for dense wide baseline image matching. International Journal of Distributed Sensor Networks (IJDSN) 12(12) (2016)

[8] Khan MA, Tembine H, Vasilakos AV. Evolutionary coalitional games:design and challenges in wireless networks. IEEE Wireless Commun. 19(2): 50-56 (2012).

[9] Gao J, Tembine H. Empathy and Berge equilibria in the Forwarding Dilemma in Relay-Enabled Networks, International Conference on Wireless Networks and Mobile Communications (WINCOM), Rabat, Morocco, Nov 2017

[10] Mikolajczyk K，Schmid C. A performance evaluation of local descriptors. IEEE T Pattern Anal 2005; 27(10): 1615–1630.

[11] Gao J, Shi F. A Rotation and Scale Invariant Approach for Dense Wide Baseline Matching. Intelligent Computing Theory - 10th International Conference, ICIC (1) 2014: 345-356.

[12] Mikolajczyk K, Schmid C. Scale & affine invariant interest point detectors. Int J Comput Vision,

2004; 60(1): 63–86.

[13] Kalal Z, Mikolajczyk K, Matas J. "Tracking-learning-detection," IEEE Transactions on Pattern Analysis and Machine Intelligence, 2012, 34(7):1409–1422.

[14] Gao J, Tembine H. Distributed Mean-Field-Type Filter for Vehicle Tracking, in American Control Conference (ACC), Seattle, USA, May 2017.

[15] Wang K, Liu Y, Gou C, Wang F Y. A multi-view learning approach to foreground detection for traffic surveillance applications, IEEE Transactions on Vehicular Technology, 2016, 65(6), 4144– 4158.

[16] Ourique CO, Biscaia EC, Pinto JC. The use of particle swarm optimization for dynamical analysis in chemical processes. Computers & Chemical Engineering. 2002; 26(12):1783-1793.

[17] Liu J, Wang QL. Application of neural network in the temperature control system of rectifying tower, Science and Technology Consulting Herald; 2007. (in Chinese)

[18] Du L, Cao JT, Li SC. Temperature control of distillation tower based on improved dynamic neural network [J]. Industry Control and Applications. 2017, 36(1): 25-31. (in Chinese)

[19] MCCULLOCH WS, PITTS W. A logical calculus of the ideas immanent in nervous activity, Bulletin of Mathematical Biophysics, 1943, 5: 115-133.

[20] HEBB DO, The Organization of Behavior: A Neuropsychological Theory[M]. Lawrence Erlbaum Associates, New Jersey, 1949.

[21] ROSENBLATT F. The perceptron: Probabilistic model for information storage and organization in the brain. Psychological Review, 1958, 65(6): 386-408.

[22] Widrow B. Adaptive `Adaline' Neuron Using Chemical `Memistors, Stanford Electronics Laboratories Technical Report, No. 1553-2 (1960)

[23] MINSKY L, SEYMOUR AP. Perceptrons: An Introduction to Computational Geometry[M]. MIT Press, Cambridge, 1969.

[24] Grossberg S. Studies of mind and brain: neural principles of learning, perception, development, cognition and motor control. Reidel Press, Boston(1982).

[25] Hopfield JJ. Neurons with graded response have collective computational properties like those of two-state neurons. Proc Natl Acad Sci, 1984, 81(10): 3088–3092.

[26] Rumelhart DE, Hinton GE, Williams RJ. Learning representations by back-propagating errors. MIT Press, Cambridge (1988).

[27] Moody J, Darken CJ. Fast Learning in Networks of Locally-tuned Processing Units. *Neural Computation*，1989，1(2): 281-294.

[28] Shi HR, Zuo F. Neural Network Modeling and Genetic Algorithm Optimization of Operating Conditions for Catalytic Distillation Synthesis of Dimethyl Carbonate. Journal of Donghua University (*Natural Sciences*),, 2006(04): 47-50. (In Chinese)

[29] Du MX, Wang YW, Zhang XZ. Optimal Design of Centrifugal Pump Based on RBF Neural Network and Genetic Algorithm. *Journal of China Three Gorges University* (*Natural Sciences*), 2020, 42(04): 88-93. (In Chinese)

[30] Lopez-Garcia TB, Coronado-Mendoza A, Dominguez-Navarro JA. Artificial neural networks in microgrids: A review. Engineering Applications of Artificial Intelligence, 2020, 95: 103894.

[31] Lecun Y, Bottou L, Bengio Y, Haffner P. Gradient-based learning applied to document recognition. Proceedings of the IEEE, 1998, 86(11): 2278-2324.

[32] Chen Q，Weigand WA. Dynamic optimization of nonlinear processes by combining neural net model with UDMC. AIChE Journal，2004，40（9）：1488-1497.

[33] Hinton GE, Salakhutdinov RR. Reducing the Dimensionality of Data with Neural Networks[J]. Science, 2006, 313(5786):504.

[34] Hinton GE, Osindero S, Teh Y. A fast learning algorithm for deep belief nets. Neural Computation, 2006, 18, 1527–1554.

[35] Gao J, Tembine H. Distributed mean-fieldtype filters for traffic networks, IEEE Transactions on Intelligent Transportation Systems. 2019; 20(2):507-521.

[36] Wang LH. Forecast Model of Short-Term Sales in E-Commerce Based on BP-AdaBoost. Computer Systems & Applications, 2021, 30(2): 260-264.(in Chinese)

[37] Martinez-Morales J, Quej-Cosgaya H, Lagunas-Jimenez J, Palacios-Hernandez E, Morales-Saldana J. Design optimization of multilayer perceptron neural network by ant colony optimization applied to engine emissions data, Science China Technological Sciences, 2019, 62(6): 1055-1064.

[38] Gao J, Tembine H. Distributionally Robust Games: Wasserstein Metric, International Joint Conference on Neural Networks (IJCNN), Rio de Janeiro, Brazil, July 2018

[39] Ng SC, Leung SH, Luk A. Fast Convergent Generalized Back-Propagation Algorithm with Constant Learning Rate. Neural Processing Letters, 1999, 9: 13-23.

[40] Ramchandran S, Rhinehart R R. A very simple structure for neural network control of

distillation , 1995, 5 (2)： 115-128.

[41] Zhao Z，Kuang G，Wang L，Zhao S, Liu J. Improvement of BP Algorithm on Network Training and Application of ANN in Catalytic Distillation Column Simulation, Chemical Engineering (China), 1998, 26(6): 44-46.

[42] Wang Y, Yao P. Advancement of simulation and optimization for thermally coupled distillation using neural network and network and genetic algorithm[J]. CIESC Journal, 2003, 54(9): 1246–1250.

[43] Shi H, Zuo F. Neural Network Modeling and GA Optimization of DMC Catalyst Distillation System, Journal of Donghua University (Natural Science), 2006, 32(4): 47-50.

[44] Liu J, Wang Q. Application of Neural Network in the Temperature Control System of Rectifying Tower, Science and Technology Consulting Herald. 2007, 1, 6-8.

[45] Ma S. Simulation of Anisole Distillation Tower Based on Artificial Neural Network. Master Thesis, Tianjin University, 2007.

[46] Konakom K, Kittisupakorn P, Saengchan A, Mujtaba IM. Optimal policy tracking of a batch reactive distillation by neural network-based model predictive control (NNMPC) strategy. World Congress on Engineering and Computer Science (WCECS)，San Francisco 2010

[47] Sharma N, Singh K. Model predictive control and neural network predictive control of TAME reactive distillation column. Chemical Engineering and Processing: Process Intensification, 2012, 59: 9-21

[48] Tun LK, Matsumoto H. Application Methods for Genetic Algorithms for the Search of Feed Positions in the Design of a Reactive Distillation Process. Procedia Computer Science, 2013, 22: 623-632

[49] Abdul Jaleel E, Aparna K.   Identification of Ethane-Ethylene Distillation Column Using Neural Network and ANFIS," 2015 Fifth International Conference on Advances in Computing and Communications (ICACC), 2015, pp. 358-361.

[50] WANG Honghai, ZHANG Yuzhen, LI Yue, et al. The orthogonal design and neural network optimization of the extractive distillation process[J]. Journal of Hebei University of Technology, 2016, 45(3):48-56.

[51] Osuolale FN, Zhang J. Thermodynamic optimization of atmospheric distillation unit[J]. Computers & Chemical Engineering, 2017, 103: 201–209.

[52] Anish KM, Pavan Kumar MV. Estimator Based Inferential Control of an Ideal Quaternary Endothermic Reactive Distillation with Feed-Forward and Recurrent Neural Networks，Chemical Product and Process Modeling, 2018, 13(2), 20170015.

[53] Yu X, Lv W, Lu S. A method for temperature control of vinyl chloride rectification based on fuzzy neural network，CN 110647186 A，2020.

Abdul Jalee et al. [49] proposed a non linear model for binary ethane-ethylene distillation column and a method of identifying distillation towers using NARX based ANFIS. Data used for identification is obtained from Daisy database. Results showed neural network model and ANFIS model were able to capture nonlinear dynamic behavior of the distillation column, and NARX based ANFIS model is more accurate with less number of iteration.

Anish et al. [52] designed a feed-forward neural network (FNN) and a layered recurrent neural network (LRNN) based two composition estimators to control the product purity for an ideal, quaternary, hypothetical, kinetically controlled, reactive distillation (RD) column. The output variables, the compositions,were estimated using the chosen tray temperatures as inputs to the estimators. The estimator based control is found to be effective for the on-spec product purity control.

Yu et al. [53] disclosed a method for the temperature control of the rectification of vinyl chloride based on a fuzzy neural network. The specific steps are as follow: firstly, collect the sensor signal and filter it, and then perform fuzzy neural network control according to the output of the signal acquisition module; finally, use BPNN to learn and modify the connection weight, Gaussian function center value and width value in the controller to improve the stability of the system.

conclusion

This article briefly describes the research progress of artificial neural networks, and summarizes its applications in distillation systems. Neural network has strong self-learning and self-organization capabilities, so it can make full use of experimental data and apply it to rectification towers that lack precise mathematical models and nonlinearities. It can avoid the heat transfer and transfer required by traditional chemical simulation methods. The calculation and basic data of quality, kinetics, thermodynamics, etc. not only have good control effect, faster response time, good dynamic performance, small overshoot, but also strong robustness and ability to adapt to changes in the control environment. With the further development of neural network technology, its application prospects in the chemical industry will be broader.